\documentclass[fleqn,12pt,twoside]{article}
\usepackage{espcrc1}

\usepackage{graphicx}
\usepackage[figuresright]{rotating}

\newcommand{\AmS}{{\protect\the\textfont2
  A\kern-.1667em\lower.5ex\hbox{M}\kern-.125emS}}

\title{Precise Determination of the Neutron Magnetic Form Factor to Higher $Q^2$}

\author{W. K. Brooks\address[JLAB]{Thomas Jefferson National Accelerator Facility, 
    12000 Jefferson Ave., \\ 
    Newport News, VA, 23606, USA}%
  \thanks{This work was supported by DOE contract DE-AC05-84ER40150 Modification No. M175, under which the Southeastern Universities Research Association (SURA) operates the Thomas Jefferson National Accelerator Facility.}
  and J. D. Lachniet\address[CMU]{Carnegie Mellon University, 5000 Forbes Ave., Pittsburgh, PA
    15213, USA}\thanks{This work was supported under DOE contract DE-FG02-87ER40315.}  for the CLAS Collaboration.
}
       
\begin{document}

\maketitle

\begin{abstract}
The neutron elastic magnetic form factor $G_M^n$ has been extracted from
quasielastic scattering from deuterium in the CEBAF Large Acceptance
Spectrometer, CLAS~\cite{CLAS}. The kinematic coverage of the measurement is
continuous over a broad range, extending from below 1~$\rm{GeV^2}$ to
nearly 5~$\rm{GeV^2}$ in four-momentum transfer squared. High precision is
achieved by employing a ratio technique in which most uncertainties
cancel, and by a simultaneous in-situ calibration of the neutron
detection efficiency, the largest correction to the data. Preliminary
results are shown with statistical errors only.
\end{abstract}

\section{INTRODUCTION}

The elastic form factors of the proton and neutron are fundamental
quantities which have been studied for decades. The dominant
features of the larger form factors $G_M^p$, $G_E^p$, and $G_M^n$ were
established in the 1960's: the dipole form $G_{dipole} = 
(1+Q^2/0.71)^{-2}$ gives a  
good description, corresponding to an exponential falloff in the spatial
densities of charge and magnetization. In the intervening decades,
obtaining higher precision measurements of these quantities has been
one thrust of the field, while new directions have also emerged,
especially over the past decade. These include precise measurements of
the neutron electric form factor~\cite{Gen}, and
extractions of the strange 
electric and magnetic form factors for the proton~\cite{G0}, as well as
time-like form factors~\cite{timelike}. In
addition to experimental progress, there has been renewed theoretical
interest on several fronts~\cite{Kees}. First, models of the nucleon
ground state can often be used to predict several of these quantities,
and it has proven to be very difficult to describe all of the modern data
simultaneously in a single model approach. Second, lattice
calculations are now becoming feasible in the few-GeV$^2$ range, and
over the next decade these calculations will become increasingly
precise. Finally, since elastic form factors are a limiting case of
the generalized parton distributions (GPDs), they can be used to
constrain GPD models. For this purpose, high precision and a large
$Q^2$ coverage is quite important~\cite{Kroll}. At present the neutron
magnetic form factor at larger $Q^2$ is known much more poorly than
the proton form factors. 

\section{THE CLAS MEASUREMENT}

The present measurement~\cite{E94-017} makes use of quasielastic scattering on
deuterium where final state protons and neutrons are detected. The
ratio of $^2\rm{H}(e,e'n)$ to $^2\rm{H}(e,e'p)$ in quasi-free kinematics is
approximately equal to the ratio of elastic scattering from the free
neutron and proton. The ratio is: 

\vspace{-0.6cm}

\begin{eqnarray}
R_D ~~ = ~~ 
{{d\sigma \over d\Omega}[^2\rm{H}(e,e'n)_{QE}] \over 
{d\sigma \over d\Omega}[^2\rm{H}(e,e'p)_{QE}]
}
~~ = ~~ a \cdot R_{free} ~~ = ~~ a \cdot { 
{(G_E^n)^2+\tau(G_M^n)^2 \over 1+\tau}
+2\tau(G_M^n)^2\tan^2({\theta\over2})
\over 
{(G_E^p)^2+\tau(G_M^p)^2 \over 1+\tau}
+2\tau(G_M^p)^2\tan^2({\theta\over2})
}
\end{eqnarray}
\vspace{-.4cm}

Using deuteron models one can accurately compute the correction
factor $a(Q^2,\theta _{pq})$,
which is nearly unity for quasielastic kinematics and higher
$Q^2$. The value of $G_M^n(Q^2)$ is then obtained from the measured value
of $R_D$ 
and the experimentally known values of $G_E^n(Q^2)$, $G_M^p(Q^2)$, and
$G_E^p(Q^2)$; this method has been used previously~\cite{Sick}. The $(e,e'n)$ and
$(e,e'p)$ reactions were measured at the same time from the same
target. Use of the ratio $R_D$ under these circumstances
reduces or eliminates several experimental uncertainties,
such as those associated with the luminosity measurement or radiative
corrections. The remaining major correction is for the detection 
efficiency of the neutron. 
\begin{figure}[t!!]
\caption{Neutron detection efficiency vs. neutron momentum in three
  CLAS subsystems. The large plot shows the efficiency from the
  forward calorimeter, the left inset compares the forward (EC) and large
  angle (LAC) calorimeters, and the right inset exhibits the efficiency in
  the time-of-flight detectors.}
\label{fig:brooks_nde}
\centerline{\includegraphics[scale=0.48,angle=-90]{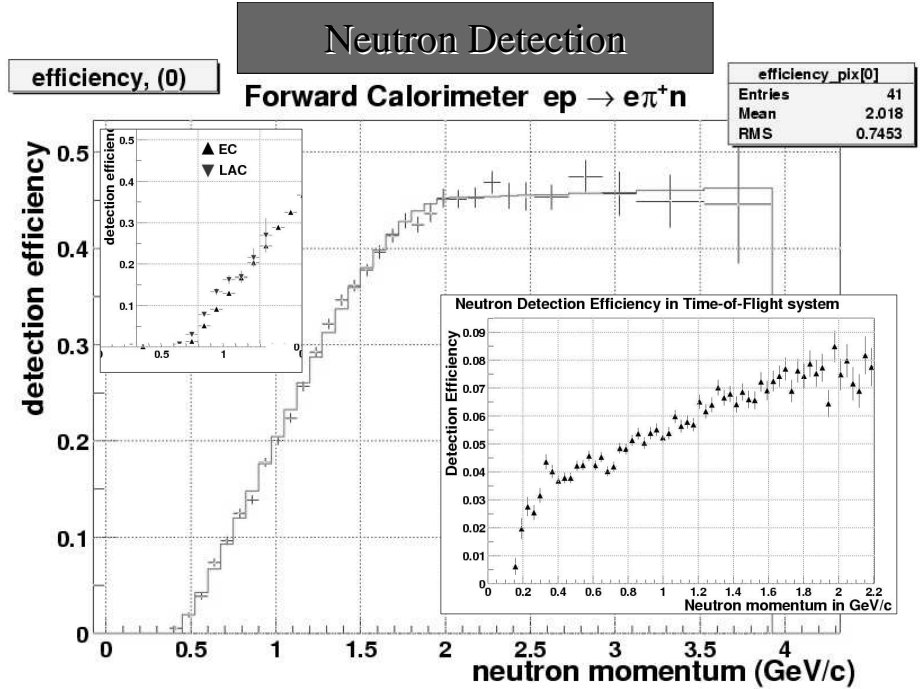}}
\vspace{-1.0cm}
\end{figure}

\subsection{Neutron Detection Efficiency}
Neutrons were measured in three CLAS scintillator-based detectors: the
forward-angle and large-angle electromagnetic shower calorimeters, and
the time-of-flight scintillators. The efficiency measurement was
performed using tagged neutrons from the $^1\rm{H}(e,e'\pi^+)X$ reaction
where the mass of the final state $M_X$ was chosen to be that of the
neutron. 

Since the precise value of the detection efficiency can vary with
time-dependent and rate-dependent quantities such as 
photomultiplier tube gain, the detection efficiency was measured
\emph{simultaneously} with the primary deuterium measurement. Two separate
targets were positioned in the beam at the same time, one for
deuterium and the other for hydrogen, separated by less than 5 cm.

A plot of the resulting neutron detection efficiency is shown in
Fig. \ref{fig:brooks_nde}. The main plot shows the results for the forward
electromagnetic shower calorimeter, while the insets show the results
for the large angle calorimeter and the time of flight
scintillators.

\subsection{Overlapping Measurements}
The CLAS extraction of $G_M^n(Q^2)$ actually consists of multiple
overlapping measurements. The time of flight scintillators cover the
full angular range of the spectrometer, while the calorimeters cover
subsets of these angles, thus $G_M^n(Q^2)$ can be obtained from two
independent measures of the neutron detection efficiency. In addition,
the experiment was carried out with two different beam energies that
had overlapping coverage in $Q^2$, so that the detection of the
protons of a given $Q^2$ took place in two different regions of the
drift chambers. As a result, essentially four measurements of
$G_M^n(Q^2)$ have been obtained from the CLAS data that potentially
could have four independent sets of systematic errors. In practice
these four measurements are consistent within the statistical errors,
suggesting that the systematic errors are well-controlled and small.

\subsection{Systematic Uncertainties}
The final evaluation of the systematic uncertainties for this
measurement has not been performed, and therefore only statistical
uncertainties are presented. It is anticipated that several
systematic uncertainties will contribute at the percent level, with a
number of others contributing at a fraction of a percent. The larger
uncertainties are expected to be due to the neutron detection
efficiency determination, the two-photon-exchange portion of the
radiative correction, uncertainties in $G_E^n(Q^2)$, $G_M^p(Q^2)$, and 
$G_E^p(Q^2)$, and suppression of inelastic background. The smaller
uncertainties are expected to be due to the proton detection
efficiency (measured by elastic scattering from the hydrogen target),
the remnant of the radiative corrections not cancelling in 
the ratio, the theoretical correction $a(Q^2)$ for quasi-free
scattering, the definition of the fiducial volume for neutrons and
protons, and a number of other small contributions. It is expected
that the ultimate uncertainties will range from two to three percent
over the full range in $Q^2$.

\begin{figure}[h!!]
\caption{Preliminary results for $G_M^n / (\mu_n G_{dipole}) $ from CLAS (see
text).}
\label{fig:brooks_results}
\centerline{\includegraphics[scale=0.5,angle=-90]{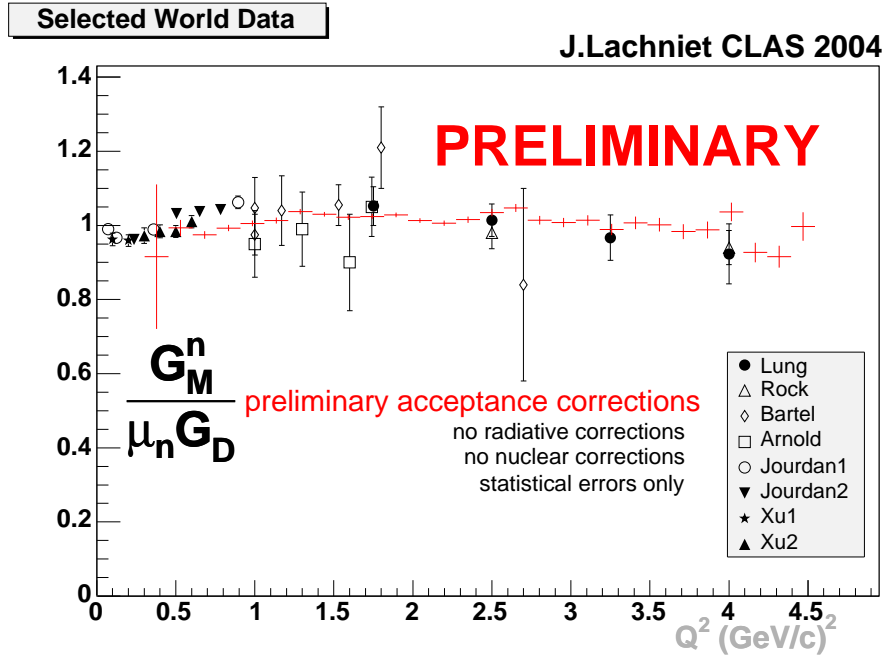}}
\vspace{-.8cm}
\end{figure}

\subsection{Preliminary Results}
The preliminary results are shown in Fig. \ref{fig:brooks_results}
together with a sample of existing data. The
error bars shown are due only to statistical uncertainties. The data
shown are the weighted averages of the four overlapping individual
measurements 
discussed above. Because these results are preliminary, it is
necessary to be cautious about the conclusions drawn, since
few-percent shifts in the results are still possible. Nonetheless, a
few features are noteworthy. First, the quality and coverage of the
data is a very substantial improvement over the existing world's data
set. Second, the dipole form appears to give a good
representation of the data over the $Q^2$ range measured, which is at
variance at higher $Q^2$ with parameterizations based on
previous data, which tend to show a more strongly
decreasing trend for $G_M^n /(\mu_n G_{dipole})$ with increasing $Q^2$.

At face value, the lowest $Q^2$ points appear to disagree with
previous high-precision data. However, these data are too preliminary to
make this conclusion. The lowest four points, unlike all others on the
plot, are not an average over multiple measurements, and they are near
the edge of the detector acceptance. Some further study is required to
establish the final centroids and uncertainties for these points.

\end{document}